# THE FCC-ee STUDY: PROGRESS AND CHALLENGES

M. Koratzinos, University of Geneva, Switzerland
S. Aumon, C. Cook, A. Doblhammer, B. Haerer, B. Holzer, R. Tomas, F. Zimmermann,
CERN, Geneva, Switzerland
U. Wienands, SLAC, USA; L. Medina, Universidad de Guanajuato, Leon, Mexico
M. Boscolo, INFN Laboratori Nazionali di Frascati, Italy; A. Bogomyagkov, D. Shatilov and E.
Levichev, BINP, Novosibirsk, Russia

*Abstract*

The FCC (Future Circular Collider) study represents a vision for the next large project in high energy physics, comprising an 80–100 km tunnel that can house a future 100TeV hadron collider. The study also includes a high luminosity $e^+e^-$ collider operating in the centre-of-mass energy range of 90–350 GeV as a possible intermediate step, the FCC-ee. The FCC-ee aims at definitive electro-weak precision measurements of the Z, W, H and top particles, and search for rare phenomena. Although FCC-ee is based on known technology, the goal performance in luminosity and energy calibration make it quite challenging. During 2014 the study went through an exploration phase. The study has now entered its second year and the aim is to produce a conceptual design report during the next three to four years. We here report on progress since the last IPAC conference.

## INTRODUCTION

The discovery of the Higgs boson with a mass of around 125 GeV has recently revived interest in circular colliders [1]. The FCC-ee study reported here was kicked off in February 2014. It is an ambitious project that aims to perform physics studies in a variety of different beam energies, the main ones being 45 GeV for precision studies of the Z boson, 80 GeV for precision studies of the W, 120 GeV for the Higgs and 175 GeV for the top quark. Other beam energies might also be needed (notably a monochromatic run at 63 GeV is being investigated). Running at different energies requires different machine configurations and parameter sets have already been published [2] to serve as a baseline. For instance, the 45 GeV running is characterised by high beam current (1.5 A) putting stringent requirements to the power couplers of the RF system. On the other hand, the 175GeV running is characterised by the need for very low emittances to mitigate the beamstrahlung problem and relatively high total RF voltage (12 GV). High momentum acceptance is also needed for high energy running (175 GeV and to a lesser extend 120 GeV), also to mitigate the beamstrahlung effects. The parameter set already published is expected to evolve as more performant solutions are being engineered. The largest potential improvement is the development of the 'crab waist' interaction region scheme, where a large crossing angle (30 mrad) and crab sextuples allow for beam-beam parameters much larger than with the head-on baseline scheme, resulting in much higher luminosity at the Z running and modest gains all the way up to the Higgs running. The past year was explicitly intended to be an exploratory phase without the aim for convergence on a specific solution. Nevertheless, work in a number of areas has resulted in multitude of papers submitted for presentation at IPAC15, which this contribution aims to summarise.

## CIVIL ENGINEERING

The civil engineering team at CERN are investigating the feasibility of an 80 km – 100 km tunnel in the Geneva basin to house the FCC machines [3]. Like any civil engineering feasibility study for a large-scale project, one of the major challenges is the management and manipulation of an enormous amount of spatial data. A feasibility study is an inherently iterative process as new data comes to light and parameters of the study evolve.

The traditional approach therefore requires significant resources in terms of time and man-power. However, to more effectively and efficiently conduct the study, CERN is employing the use of a specially designed interactive tool, containing a 3D geological model of the Geneva basin. The Tunnel Optimisation Tool (TOT) has been developed and is based on commercially available Geographical Information System (GIS) software.

A tunnel design is uploaded to TOT and positioned within the geology of the Geneva basin. The tool then outputs key information for a given setup including the geology intersected by the tunnel, the shaft depths, interaction with the built-environment, including buildings and geothermal boreholes and interaction with the environment including environmentally protected areas.

Another challenge for the study has been making a quantitative comparison between different solutions. Given the high number of variables and multiple objectives for optimisation (minimise shaft depths, minimise tunnel length in limestone, minimise length and slope of injection tunnels from the LHC, etc.), some form of detailed analysis is required. Comparison also needs to not only be made between two positions of the same tunnel but also between two tunnels of different circumference.

A list of factors related to the cost/risk of construction of each element of the FCC project (tunnel, shafts and caverns) have been compiled which are based on engineering experience of tunnelling projects. The data from TOT is extracted and multiplied by the relevant factor, giving a total cost/risk value to any given solution.

An alternative method of analysis is also under investigation. This is the use of an optimisation algorithm,

ROXIE, used previously for the optimisation of magnet design at CERN. Based on a set of pre-defined objective functions, it may be possible for TOT to search for the optimal position of the FCC tunnel mathematically.

Early results show that when comparing tunnels of 80km, 87 km, 93 km and 100 km, there is an inevitable linear increase in the cost/risk of any solution which relates simply to the extra length of tunnel required. This linear increase in cost/risk has been removed from the results to leave a measure of how suitable the circumferences are for the geology and terrain of the FCC study area, or how well they 'fit' into the Geneva basin. However, the results show that an increase in cost/risk still remains as the circumference is increased from 80 – 100km. This is a consequence of the 87 km, 93 km and 100 km tunnels extending further north than the 80 km and therefore travelling under a deeper section of Lake Geneva. This leads to increased tunnel depth along the circumference of the tunnel and consequently increased shaft depths. The 100 km tunnel is also penalised in this analysis for part of the tunnel intersecting with the Jura limestone, something which the other three options are able to avoid.

## BEAM OPTICS

The FCC-ee study is based on racetrack geometry with a circumference of about 100 km and foresees running at four different centre-of-mass energies to allow precision measurements at the 45, 80, 120 and 175 GeV beam energy. For each energy the beam parameters depend crucially on the synchrotron light emission and the lattice has to be optimised to provide the emittance target values of the baseline design [2]. A first lattice has been designed and currently been optimised [4]. The basic cell has been chosen for 175 GeV operation. It combines four dipole magnets and two main quadrupoles in a 50 m long FODO cell. A phase advance per cell of 90/60 degrees is considered as a first iteration for 175GeV. The layout of the beam optics follows the typical pattern:

- A standard FODO cell in the arcs to define the beam emittance for the four energies.
- Matching sections to provide smooth modification of the optics towards the straight sections.
- Dispersion suppressors to establish dispersion free straight sections for installation of the 4 experiments and the RF stations.
- Four interaction regions to achieve the mini-beta values including a feasible beam separation scheme as well as a local chromaticity correction.

For 45 GeV running, to increase the emittance to the values required in [2], the effective cell length is increased by rearranging the quadrupole focusing scheme, increasing the FODO cell length from 50 m to up to 300 m.

In order to achieve the required luminosities the vertical beta function at the interaction point (IP) has to have the very small value of β* =1 mm. This has considerable impact on the layout of the interaction region. A considerable crossing angle is needed to avoid parasitic bunch crossings on either side of the IP, and at the same time the synchrotron radiation emitted next to the detector has to be kept low enough. The large excursions of the amplitude function combined with the small β* values impose serious tolerances for magnet alignment and coupling compensation schemes. Last but not least, the chromaticity compensation scheme needed has to combine an optimised arc sextupole distribution scheme with a local correction in order to guarantee a sufficiently large off-momentum dynamic aperture of the machine extending up to Δp/p≈±2%. Between the final mini-beta telescope, which is based on two doublets and a matching section to obtain a smooth transition to the arc optics, a sequence is embedded which provides locally sufficient dispersion and sextupole strength to compensate a large fraction of the interaction region (IR) chromaticity. The combination of the local sextupole scheme in the IR and an optimised sextupole distribution in the arc design, yields continuous stable linear optics between -2% and 2%. This is a very positive first step towards full beam stability in the $dp/p = \pm 2\%$ range. The combination of the local sextupole scheme in the IR and an optimised sextupole distribution in the arc design, yields a momentum acceptance of ~2%. Therefore one of the most critical requirements of the design seems to be within reach.

## INTERACTION REGION

The interaction region of FCC-ee is particularly challenging as luminosity optimisation ought to be performed while satisfying a number of constraints relating to beamstrahlung, synchrotron radiation loss around the IR, acceptable final focus quadrupole design, etc. The minimum distance from IP to the face of the first quadrupole is chosen to be L* = 2 m which at the present moment looks like a good compromise between beam dynamics and detector constraints. A particularly promising approach is the introduction of a crab waist scheme [5] with a crossing angle of 30 mrad. The advantages of this approach is high luminosity (mainly at 45 and 80 GeV running) and natural separation of bunches compared to a small or no crossing angle approach. The challenges of this scheme are as follows: the small values of IP beta functions, essential for high luminosity, produce large chromaticity, which should be compensated as locally as possible in order to minimize excitation of nonlinear chromaticity; synchrotron radiation power loss should be significantly smaller than in the arcs; synchrotron radiation at high beam energies will produce photons with high critical energy which increases the detector background; finally, small beta functions at the IP enhance effects of nonlinear dynamics, decreasing dynamic aperture and energy acceptance of the ring, both essential for good performance.

The study of the scheme is quite advanced. A geometrical layout exists and synchrotron radiation energy loss requirements are the subject of optimization. Shifting sextupoles in phase with respect to final focus quadrupoles proves to be an efficient method to control second order chromaticity of the betatron phase advances. The introduction of an additional sextupole four times weaker than the main sextupoles at the place where beta is small (and the second order chromaticity is large) gives a useful knob to control third order chromaticity. The energy acceptance for linear stable optics of one quarter of the ring is [–3.9%; +1.9%].

## INTERACTION REGION LOSSES

The interaction regions of high-luminosity colliders like the FCC-ee require well balanced designs, which provide both for very high luminosity and at the same time keep backgrounds and radiation at tolerable levels. Losses at the interaction region can be divided in two main sources: losses of beam particles and synchrotron radiation. Regarding losses of beam particles, we plan to adapt a tool first developed for Flavour Factories [6]. Regarding synchrotron radiation, a set of flexible tools targeted at providing a first evaluation of losses in the IR is being developed – MIDSim [7]. This tool can read the machine description, visualise geometry and perform detailed simulation of the passage of particles through materials. The application of the tool to the early design of the IP region with the 30 mrad crossing angle reveals synchrotron radiation (SR) power higher than 4 MW per IP and critical energies at 175 GeV running of 4.4 MeV, so further iterations would be needed for this very challenging area of the FCC-ee study.

## DYNAMIC APERTURE

Dynamic aperture (DA) studies have been conducted [8] by particle tracking using MAD-X and PTC at the 175 GeV operation energy. Two different schemes for the IR have been considered: the first follows a crab waist approach [5] while the second was obtained from the former by decreasing the crossing angle from 30 to 11 mrad, removing the crab and matching sections, and scaling its total length to 700 m as for the crab waist approach. The change of crossing angle decreases the synchrotron radiation power going into the detector region from 4.4 to 0.4 MW.

The DA obtained for both schemes up to now is still below what is needed and optimisation is still ongoing. A refined matching between the IR and the ring optics –with special care of the Montague W functions–, the choice of an appropriate working point, the reduction of L* to reduce the chromaticity, and the correction of higher order chromaticities, may improve the properties of both designs.

## ENERGY CALIBRATION

Accurate energy determination is a fundamental ingredient of precise electroweak measurements. A strategy based on resonant depolarization measurements drawing from the LEP experience is being developed [9]. Resonant depolarization measurements ought to be performed at a rate of a ~5 per hour on 250 non-colliding, dedicated, bunches. Both electrons and positrons should be measured. To get over very large polarization times an asymmetric wiggler scheme needs to be implemented. These wigglers consume 15-30% of the SR power of the machine, but need only be used for about 50-100 minutes after filling up the machine. Polarization at W energies (80 GeV per beam) seem possible, in contrast to LEP. The first simulation results are encouraging and show that polarization levels of a non-perfect machine (using as a typical misalignment of quadrupoles $y\delta_{RMS}^{y} = 200 \mu m$) can be restored by careful use of correctors and harmonic bumps.

## STAGING SCENARIOS

Although the study is still at an early stage and not all information is available, a first attempt has been made to examine possible staging scenarios [10]. An obvious candidate for staging is the RF system, mainly due to conflicting requirements of high beam current operation at the Z and high voltage at the top running (175 GeV). Reconciling the low angle IR design to the high angle one (which is suitable for crab waist operation) can offer not only a better understanding of the issues but also a path for optimisation of the crossing angle. Such an optimization is multivariate and not yet complete.

## CONCLUSIONS

The FCC-ee study enters its second year, following a period of exploration of different approaches and strategies. An optics design fulfilling the requirement for emittance has been designed. Two approaches have been pursued for the interaction region, with the crab waist approach being the one where most effort was concentrated and which apparently gives the best performance. Optics momentum acceptance is close to the very challenging requirement of 2%. Dynamic aperture studies have started and progress has been made, although the dynamic aperture has not yet reached the required level. Synchrotron radiation losses around the interaction point are currently high, and an effort has started to have them reduced to tolerable levels. Another area of intense activity is again the interaction region and more specifically the magnetic design around the IP in the presence of the detector solenoid, the final focus quadruples, and screening and compensation solenoids. The study looks forward to its second phase which aims to produce a conceptual design report in about three years from now.